# Optimization of Enzymatic Logic Gates and Networks for Noise Reduction and Stability

Mary A. Arugula, Jan Halámek, Evgeny Katz, Dmitriy Melnikov,
Marcos Pita, Vladimir Privman, and Guinevere Strack

*Department of Chemistry and Biomolecular Science and Department of Physics*
*Clarkson University, Potsdam, NY 13699, USA*

**Abstract**

Biochemical computing attempts to process information with biomolecules and biological objects. In this work we review our results on analysis and optimization of single biochemical logic gates based on enzymatic reactions, and a network of three gates, for reduction of the "analog" noise buildup. For a single gate, optimization is achieved by analyzing the enzymatic reactions within a framework of kinetic equations. We demonstrate that using co-substrates with much smaller affinities than the primary substrate, a negligible increase in the noise output from the logic gate is obtained as compared to the input noise. A network of enzymatic gates is analyzed by varying selective inputs and fitting standardized few-parameters response functions assumed for each gate. This allows probing of the individual gate quality but primarily yields information on the relative contribution of the gates to noise amplification. The derived information is then used to modify experimental single gate and network systems to operate them in a regime of reduced analog noise amplification.

## 1. Introduction

As an information processing network becomes larger and the information is processed in greater quantities and at higher levels of complexity, noise inevitably builds up and can ultimately degrade the useful "signal" which is the intended result of the logic processing or computation. One then has to develop approaches to achieve what is known as "fault-tolerant" information processing that involves noise control and suppression.

Presently we are aware of three fault-tolerant information processing paradigms. The first one is the familiar analog/digital electronics paradigm of the silicon-chip technology in modern computers. We know how to design such devices, and they have been successfully built. Living organisms are the second paradigm. While we obviously know that this paradigm leads to a functional scalability (one notable example is the "robustness" of many complex processes in cell functions), we do not yet fully understand it, though significant strides have been made in systems biology to explore the structure and functioning of biological "networks." The third, recent paradigm, involves massive parallelism — quantum-mechanical (quantum computing) or ensemble (variants of DNA computing).

Biochemical computing — in our experiments and modeling work, based on enzymatic reactions [1-8] — attempts to process information with biomolecules and biological objects [9], primarily using the analog/digital information processing paradigm of ordinary electronics. Indeed, most biochemical computing studies attempt to realize and, recently, network "gates" that mimic Boolean digital logic.

Biomolecular systems offer the advantage of specificity (i.e., of being selective) in their chemical functions and therefore usable in complex "chemical soup" environments. Many processes in a living cell are controlled (catalyzed) by specific enzymes and create a complex network of interconnected biochemical reactions. This means that in principle, enzyme-based information processing units can be made highly scalable giving rise to artificial biocomputing networks [6,7,10] performing various logic functions and mimicking natural biochemical pathways [10]. Compared with electronic counterparts, biomolecular computing systems also have the advantage of being able to process information in the form of chemical inputs directly from biological systems. This is important for interfacing of the resulting bioelectronic devices with living organisms, for potential applications. Computational networks that solely involve biochemical processes [6,11] are being researched for new technological capabilities. The ultimate goal has been removing the batteries and/or generally reducing the need for inorganic leads and electrical power supply, at those stages of information processing that occur during biomedical testing, in implantable devices, and other fast decision making steps in (bio)medical applications.

Recent experimental advances in enzyme-based biocomputing have included not just experimental demonstrations of several single Boolean gates such as AND, OR, XOR, INHIB, etc. [1-8], but also networking of several (at present, up to 3-4) gates [6,11]. Similar logic operations were also realized using non-biological chemical systems [12].

The increasing complexity of enzyme-based logic networks presents interesting challenges. These include: analog noise suppression through gate and network optimization; control of digital noise through network redundancy; and development of biochemical filters, rectifies and in general universal logic gates and network elements. Within the analog/digital information processing paradigm, error buildup can be suppressed by gate optimization for reduction and control of the analog noise amplification [5], as well as by network design and/or network topology [6]. For larger networks, another, "digital" [5-7] mechanism of noise amplification emerges which can be controlled by utilizing redundancy in network design and requires truly digital information processing with appropriate network elements for filtering, rectification, etc.

The present sizes of the biochemical computing networks should allow exploration of design and optimization issues related to suppression of the "analog" noise amplification. The work on this topic was initiated in our recent publications [5-8]. Here we survey the issues of the analog noise suppression in biocomputing. Section 2 addresses the functioning and optimization of a single Boolean logic gate. Section 3 considers optimization at the level of a network of several connected gates.

## 2. Optimization of a single biochemical Boolean logic gate

### 2.1. Theoretical model

We explore how scalability paradigms for complex information processing can be adapted from ordinary electronics to biochemical logic. Specifically, individual biocomputing gates will be first analyzed for decreasing the level of "analog" noise amplification [5,8].

We consider a specific example of a gate with two chemicals as inputs into an enzyme-catalyzed reaction. One of these is the co-substrate (to be specified later, since we consider two options), the concentration of which will be denoted $I_1$. Another is the substrate, hydrogen peroxide ($H_2O_2$) which is the second logic input, $I_2 = [H_2O_2](t=0)$. The reaction is catalyzed by the enzyme horseradish peroxidase (HRP) at the initial concentration $E(t=0)$. As described in [8], for the AND logic-gate function parameterization we use a simplified (approximate) kinetic description with of two irreversible steps, assuming a single intermediate complex, $C$,

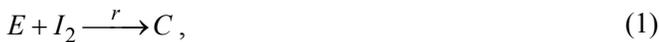
(1)

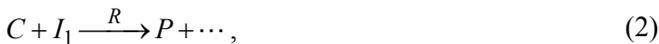
(2)

where $P$ is the concentration of one of the two products (the oxidized co-substrate) which is optically detected as the logic-gate output, while $r$ and $R$ are the effective rate constants for the reactions. Note that the labeling of the logic inputs is arbitrary, dependent on application, and not correlated with the order of their intake in the enzymatic reactions Eq. (1-2).

The corresponding rate equations are

$$\frac{dI_2(t)}{dt} = -r\bigl[E(0) - C(t)\bigr]I_2(t), \qquad (3)$$

$$\frac{dC(t)}{dt} = r\bigl[E(0) - C(t)\bigr]I_2(t) - RC(t)I_1(t), \qquad (4)$$

$$\frac{dI_1(t)}{dt} = -RC(t)I_1(t), \qquad (5)$$

with the relations $E(t) = E(0) - C(t)$, $P(t) = I_1(0) - I_1(t)$, the second of which yields the product, $P(t)$, concentration once Eq. (3-5) are solved for $I_{1,2}(t), C(t)$. Here we assume that initially $P(0), C(0) = 0$. The initial $I_{1,2}(0)$ values depend on the environment in which the enzymatic gate is used and on the "logic state" (see below), whereas the gate "activity" can be adjusted by our selection of $E(0)$. Thus, even with the present simplified modeling of the HRP enzymatic kinetics, we have several parameters to deal with during the fitting of the gate function and its optimization.

For the AND gate function, we set Boolean 0 (logic-0) as zero (initial) concentration of each of the input chemicals, $I_{1,2}(0)$, and the output, $P(t = t^{\text{gate}})$, at the particular "gate time" $t = t^{\text{gate}}$. The Boolean 1 (logic-1) inputs correspond to reference concentrations of $I_1^{\text{gate}}$ and $I_2^{\text{gate}}$ at time 0 which together with the gate time $t^{\text{gate}}$ in our case were selected as experimentally convenient values, but in applications will be set by the gate environment or by the preceding gates in a logic circuit. Finally, the logic-1 output corresponds to the value $P^{\text{gate}}$ at time $t^{\text{gate}}$ set by the gate itself and therefore generally cannot be adjusted.

Ideally, our logic gate should only have chemical concentrations at logic-0 or 1 values. However, due to noise in the system, concentrations not precisely corresponding to 0 or 1 are also possible. Let us define the (dimensionless) "logic" variables

$$x = \frac{I_1(0)}{I_1^{\text{gate}}}, \quad y = \frac{I_2(0)}{I_2^{\text{gate}}}, \quad z = \frac{P(t^{\text{gate}})}{P^{\text{gate}}}, \qquad (6)$$

in terms of which we can then consider the gate response function

$$z = F(x, y) = F(x, y; E(0), r, R, t^{\text{gate}}; \ldots). \qquad (7)$$

This function can then be studied for general $x, y, z$ ranging from 0 to 1 (and to values somewhat larger than 1). The second expression in Eq. (7) emphasizes that the gate response function also depends on adjustable parameters, such as $E(0), r, R, t^{\text{gate}}$, that can in principle be varied to improve the gate performance, as well as on parameters (marked by "…") which are externally fixed.

Since the noise in general leads to some spread in the input variables around their logic-0 and 1 values, the output logic variable $z$ also becomes not precisely determined. If we assume that the magnitudes of the deviations $\delta x$ and $\delta y$ are small and comparable to each other, then the resulting deviation, $\delta z$, for *smoothly varying* gate functions can be estimated as $\delta z \sim |\nabla F| \, \delta x$, where the $|\nabla F|$ is the magnitude of the gradient vector of the gate function at the appropriate logic point (00, 01, 10, 11). This suggests that depending on the value of the largest of the four logic-input point gradients, the gate can amplify the noise level, $\delta z > \delta x, \delta y$, suppress it, $\delta z < \delta x, \delta y$, or keep it approximately constant, $\delta z \approx \delta x, \delta y$. The best-case scenario is, of course, the suppression of noise but this is only possible when the response function has a "sigmoid" shape in both $x$ and $y$ variables [5] which has not been achieved for enzyme-based gates thus far. Since our enzyme (HRP) is expected not to possess the "self-promoter" (sigmoid) response for the inputs, the best we could hope for in our biochemical system is signal propagation without noise amplification, i.e., $\delta z \approx \delta x, \delta y$.

Following [5], to estimate noise amplification we study the width of the output signal distribution $\sigma_{out} = \sigma_z$ as a function of the width of the input noise distributions assumed equal for simplicity, $\sigma_{in} = \sigma_x = \sigma_y = 0.1$. Furthermore, we will assume uncorrelated, Gaussian input noise distributions, $G_{0\text{ or }1}(x)$, with half-Gaussian for $x$ at logic-0 and full Gaussian at logic-1, and similarly for $y$. The output, $z$, distribution width $\sigma_{out}$ is then estimated by calculating $\sigma_{out}^2 = \langle z^2 \rangle - \langle z \rangle^2$ (0) for logic-1 (0) points, with the moments such as $\langle z^2 \rangle$ of the gate response function $z = F(x,y)$ computed with respect to the product input distribution $G_{0\text{ or }1}(x) G_{0\text{ or }1}(y)$.

This computation yields spread of the output signal near the respective logic value 0 or 1 for the four logic input combinations 00, 01, 10, and 11. In general, one would want to have the maximum of these spreads, $\sigma_{out}^{max}$, to be as small as possible. In fact, for network scalability the actual value of the noise spread is not as important as the degree of noise *amplification* at each gate, measured by $\sigma_{out}^{max}/\sigma_{in}$. As described earlier, the fact that the "gate machinery" enzyme (HRP in our case) is not expected to have self-promoter input(s) property ("sigmoid" gate-function shape [5]) limits $\sigma_{out}^{max}/\sigma_{in}$ to values slightly over or equal to 1.

Using the above approach, in the next subsection review our identification [8] of a regime of functioning of a single enzyme with an unusual shape of the logic-variable response surface (see below) which can yield logic gates with practically no analog noise amplification. It turns out that the rate constant $r$, see Eq. (1-2), is very large [13]. On the other hand, the rate constant $R$ describing the oxidation of the co-substrate $I_1$ can be varied in a broad range as it depends on the choice of this input chemical [14]. We demonstrated [8] that the ratio between $r$ and $R$ dramatically affects the degree of amplification of the analog noise generated by this gate, and, for sufficiently large $r/R$, yields a new response-surface shape with desirable low noise-amplification properties. We used 2,2'-azino-*bis*(3-ethylbenzthiazoline-6-sulphonic acid) (ABTS) as the "fast" co-substrate and $K_4Fe(CN)_6$ (ferrocyanide) as the "slow" co-substrate to experimentally confirm these theoretical observations; see Section 2.2.

## 2.2. Single gate optimization

The experimental response function for the gate with ABTS is shown in Fig. 1. In order to perform numerical optimization of the gate, we first fitted the experimental data by using Eq. (3-5) at the reaction time $t^{gate} = 60\,\text{sec}$. This yields estimates of the rate constants $r$ and $R$. The resulting fitted response surface is also presented in Fig. 1. The fitted rate $r = 18\,\mu M^{-1} s^{-1}$ for the reaction step involving $H_2O_2$ is large [8,13]. The reaction step involving ABTS has a somewhat slower rate $R = 5\,\mu M^{-1} s^{-1}$ which is nevertheless also quite fast and comparable to $r$ [8,13]. The results for the output noise distribution width are shown in Fig. 1. We conclude that the "figure of merit" $\sigma_{out}^{max}/\sigma_{in}$ cannot be made smaller than $\approx 3$ for the scanned parameter ranges. This means that this ABTS AND gate significantly amplifies analog noise and it is not suitable for utilization in information-processing networks.

The above conclusion is not surprising and was alluded to in the earlier work [5]. Indeed, the shape of the response function for the AND gate with ABTS (Fig. 1) is rather smooth, and therefore, one could work directly with gradients at the logic points rather than with the noise distributions. However, as shown in the next section, "balancing of gradients" at the logic points in order to make the largest of them as small as possible, can yield at best values around 1.2. Even these values are not easy to achieve unless the enzyme concentration, for instance, is varied over a large range of a couple of orders of magnitude, which is not experimentally feasible.

However, HRP is known to take on a variety of co-substrates [14]. This offers an opportunity to have a large variation in the rate $R$. Our numerical studies of the desired rate ranges based on Eq. (3-5) yielded an interesting conclusion that one can achieve values of $\sigma_{out}^{max}/\sigma_{in} \approx 1$ in the regime of large imbalance of the reaction rates, $R \ll r$.

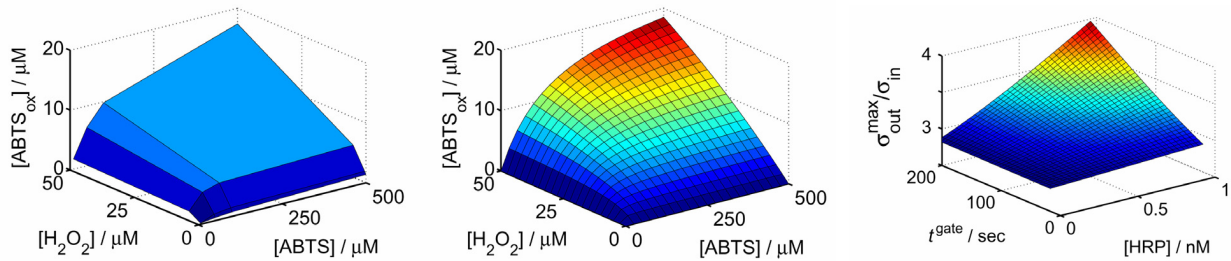

**Figure 1.** Measured (left panel) and numerically fitted (center panel) response surface for the enzymatic logic gate with ABTS as one of the inputs. Right panel: Surface plot of the gate function quality measure, $\sigma_{out}^{max}/\sigma_{in}$, as a function of the enzyme concentration and reaction time. Experimental details can be found in [8].

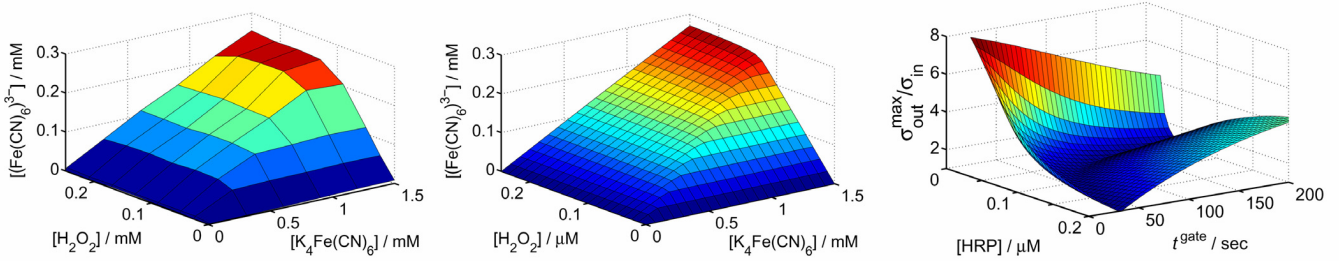

**Figure 2.** Similar to Fig. 1, measured (left panel) and numerically fitted (center panel) response surface for the enzymatic logic gate with ferrocyanide as one of the inputs. Right panel: Surface plot of the gate function quality measure, $\sigma_{\text{out}}^{\max}/\sigma_{\text{in}}$, as a function of the enzyme concentration and reaction time. (See [8] for details.)

To verify this prediction, we replaced ABTS with a "slower" co-substrate: ferrocyanide. The experimental and fitted response functions for the AND gate with ferrocyanide input, are shown in Fig. 2. One can see that these surfaces can be roughly represented by two intersecting planes. The reaction rates obtained from the fitting of the experimental data are $r = 17\,\mu M^{-1} s^{-1}$ and $R = 32 \cdot 10^{-3}\,\mu M^{-1} s^{-1}$ [8,14]. Note that the value of $R$ is about 170 times smaller than that for ABTS. From Fig. 2 one can see that the width of the output noise distribution shown achieves a marked minimum at which $\sigma_{\text{out}}^{\max}/\sigma_{\text{in}} \approx 1$ for properly selected values of the two parameters $E(0)$, $t^{\text{gate}} = E^{\min}, t^{\min}$, and that $E^{\min}(t^{\min})$ is a monotonically decreasing function. Figure 2 also suggests that our experimentally convenient but otherwise randomly selected values of $E(0) = 0.5\,\mu M$ and $t^{\text{gate}} = 60\,\text{sec}$ correspond to $\sigma_{\text{out}}^{\max}/\sigma_{\text{in}} \approx 2$. This value is already better than for ABTS but still requires optimization before the AND gate can be used as part of a network.

As pointed out in Section 2.1, some parameters of the gate response function might be fixed by the gate's surroundings (network). Let us consider, for instance, $I_2^{\text{gate}}$ — the concentration of the input $H_2O_2$. If variation of this parameter *is possible* in a particular application, then its adjustment required to achieve optimal gate functioning (with all the other parameters fixed) is also quite reasonable, as shown in Fig. 3. Reduction of the $I_2^{\text{gate}}$ concentration from our original value of 250 µM to 150 µM yields $\sigma_{\text{out}}^{\max}/\sigma_{\text{in}}$ very close to 1, resulting in noise amplification of at most 5%.

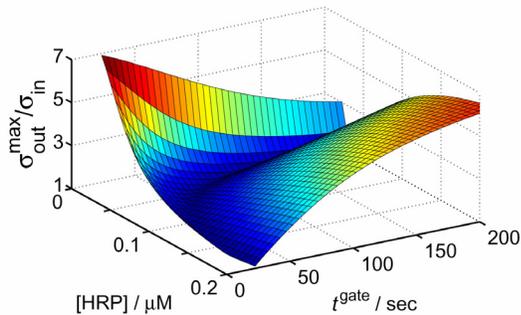

**Figure 3.** Surface plot [8] of the gate function quality measure, $\sigma_{\text{out}}^{\max}/\sigma_{\text{in}}$, as a function of the enzyme concentration and reaction time, similar to Fig. 2, but for the initial $H_2O_2$ concentration $I_2^{\text{gate}} = 150\,\mu M$.

## 3. Network optimization and design

As mentioned in Section 2.1, optimization of gates one at a time is not always a straightforward procedure since the gate response function, Eq. (7), depends parametrically on several biochemical quantities some of which are difficult to change in a broad interval. Therefore, we have also explored [6] optimization of a *network* of enzymatic reactions as a whole. In this section, we illustrate our approach on the example of the network of three AND gates, shown in Fig. 4.

We seek a simple, few parameter *modular* description of the network elements that will allow us to "tweak" the relative gate activities in the network to improve its stability. For a one-variable convex function, variation of a gate's activity rebalances the slopes near 0 and 1 with respect to each other: we need at least one phenomenological parameter to describe this shape in the simplest way possible. A convenient fit function is $x(1+a)/(x+a)$. For a single AND gate, such as the one shown in Figs. 1-2, we will thus use the product form, $F(x,y) = xy(1+a)(1+b)/(x+a)(y+b)$ with two adjustable parameters, $0 < a, b \leq \infty$.

Technically, we expect that if this proposed approximate description is at all accurate for a given gate, then the parameters $a(t_{\max}; E; k_\alpha, k_\beta, \ldots; I_\gamma, I_\delta, \ldots)$ and $b(t_{\max}; E; k_\alpha, k_\beta, \ldots; I_\gamma, I_\delta, \ldots)$ will be functions of the adjustable variables such as concentrations ($E, I_\gamma, I_\delta, \ldots$), rate constants ($k_\alpha, k_\beta, \ldots$), reaction time ($t_{\max}$), etc.

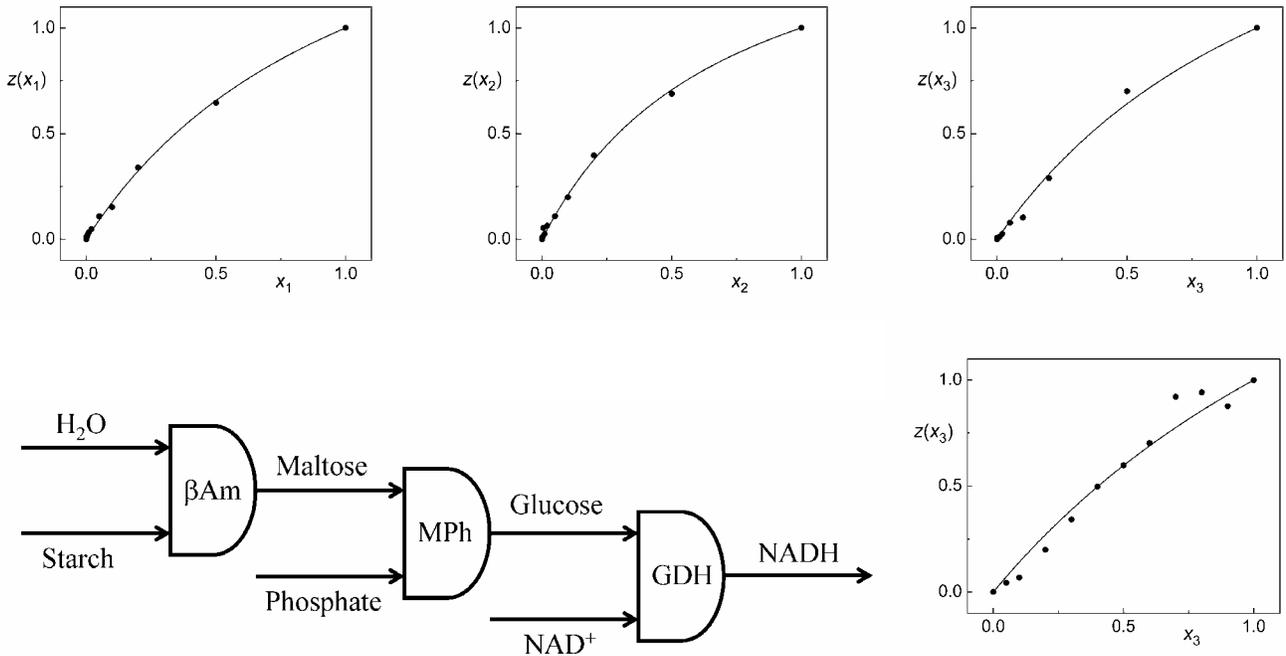

**Figure 4.** Bottom-left panel: Network of AND gates carried out by β-amylase (βAm), followed by action of maltose phosphorylase (MPh), and then by glucose dehydrogenase (GDH), with the output detected optically. Top panels: data fits [6] of the measured outputs $z(x_3)$, $z(x_2)$, $z(x_1)$, with all the other inputs held at their logic-1 values. Bottom-right panel: Example of a data fit [6] for an experiment with the optimized choice of parameters.

However, without detailed rate-equation kinetic modeling, this dependence is not known, and we cannot verify that this functional form provides a good approximation for the actual $(x,y)$-dependence of the gate-function response surface. We will not attempt such a modeling for each gate in the network. Instead, we use the approximate description to derive information on the relative network functioning by selective probes of responses to inputs, and we attempt to optimize the overall performance.

The minimum of the largest of the four gradient values of the fitting function is achieved at the optimal values $a_{\text{optimal}} = b_{\text{optimal}} = 1/(\sqrt[4]{2} - 1) \approx 5.3$. We also introduce the quantities $A = a/(1+a)$, $B = b/(1+b)$, both with optimal values $A, B_{\text{optimal}} = 2^{-1/4} \approx 0.84$. The range for these parameters is $0 < A, B \leq 1$. We note that the actual gradient values in the optimal, symmetric ($a = b$) case are $\sqrt{2}a/(1+a) = \sqrt{2}A$ for the logic-11, $1/A$ ($1/B$) for 01 (10), and 0 for 00. With the optimal parameter selections, the gradients at the three logic points 01, 10, 11, are $\sqrt[4]{2} \approx 1.189$. This means that the optimized gate-functions still somewhat amplify analog noise, by approximately 19% per processing step. This property of the convex gate response functions [5] was discussed earlier.

We now turn our attention to the specific 3-gate network shown in Fig. 4. We follow the convention of numbering the gates in their sequence, counting from the output. Suppose that we set the inputs $x_{2,3} = 1$, and measure the function $z(x_1)$. In our experiments (Fig. 4), described in detail in [6], we varied the concentration of input NAD$^+$ in the range corresponding to $x_1 \in [0,1]$. Then the resulting data should be fitted to the functional form $z(x_1) = x_1/[(1 - A_1)x_1 + A_1]$, where for brevity we omit all the fixed arguments. For variation of the output as a function of the other two inputs, phosphate (gate 2) and starch (gate 3), we get $z(x_2) = x_2/[(1 - A_2 B_1)x_2 + A_2 B_1]$ and $z(x_3) = x_3/[(1 - A_3 B_2 B_1)x_3 + A_3 B_2 B_1]$. Interestingly, in terms of the variables $A$ and $B$, not only the input $x_1$ dependence, but also the input $x_{2,3}$ dependences require single-parameter fits (see Fig. 4). Thus, variation of $x_1$ provides information on one of the phenomenological parameters, $a_1$, of gate 1. Variation of $x_{2,3}$ does not actually lead to a more complicated several-parameter data fit, even though the signal being varied, goes through more than one gate before affecting the output. Instead, we get information on a combination of parameters from more than one gate.

Our first set of data (Fig. 4) was collected with the experimentally convenient but otherwise initially randomly selected values for the adjustable "gate machinery" parameters which we will limit here to the initial enzyme concentrations, $E_{1,2,3}$, for definiteness. The collected data were rescaled into the logic variable ranges between 0 and 1 and fitted according to the single-parameter equations. We note that since not all the inputs of all the gates are varied to probe the response of the final output, we do not get all the 6 phenomenological fit parameters.

**Table 1.** Gate-function "quality measures" derived from the data fits for the initially selected parameter set and for the *modified* set (after optimization of the network functioning). (See [6] for details.)

| Input Varied | Quality Measures[a,b] (the best possible value of these quantities is $\sqrt[4]{2} \approx 1.19$) | Slope Signal | Fixed-Time Signal | Slope Signal | Fixed-Time Signal | Gates Involved |
|---|---|---|---|---|---|---|
| | | **Initial experiment** | | **Improved (optimized) experiment** | | |
| $x_1$ | $\max[\sqrt{2}A_1, 1/A_1]$ | 1.52[a] | 1.92[a] | 1.23 | 1.41 | 1 |
| $x_2$ | $\max[\sqrt{2}\sqrt{A_2B_1}, 1/\sqrt{A_2B_1}]$ | 2.04[a] | 1.56[a] | 1.41 | 1.41 | 1, 2 |
| $x_3$ | $\max[\sqrt{2}\sqrt[3]{A_3B_2B_1}, 1/\sqrt[3]{A_3B_2B_1}]$ | 1.26[a] | 1.21[a] | 1.21[b] | 1.24 | 1, 2, 3 |

[a] For the initial experiment, all the max[…] values were realized as inverses, rather than as products with $\sqrt{2}$.

[b] For the optimized experiment, this single value was realized as an inverse; all the other values were $\sqrt{2}$….

In fact, we only get one parameter and two additional combinations of parameters. Thus, we can only draw a limited set of conclusions regarding the network noisiness. For the following discussion, the data were recast (see Table 1) in terms of the geometric means of the parameters that are known only as combinations (as products). Furthermore, since in the optimal-value case the gradients at the non-00 logic-points, are actually $1/A$ (or $1/B$) and $\sqrt{2}A$ (or $\sqrt{2}B$), we took the maxima of these quantities to compare with the optimal (the smallest possible) value of the gradients, $\sqrt[4]{2} \approx 1.19$.

The following semi-quantitative conclusions follow from considering the data. Both the time-dependence-slope based and the value (at $t_{\max}$) based signal definitions (detailed in [6]) give qualitatively similar results. Gate 3 seems to be the least noisy, whereas the larger "noise amplification measure" values that involve the other two gates should be attributed to gate 1 which contributes to both measures and is thus the primary candidate for parameter modification. In fact, the maximal values in Table 1 were *all* realized with the $1/A$ or $1/B$ type value combinations, rather than the combinations involving $\sqrt{2}A(=\sqrt{2}a/(1+a))$ or $\sqrt{2}B$ type expressions. This suggests that the gradients are generally larger at logic 01 points and 10 points, as compared to logic 11 points. One way to decrease noise amplification in our network is thus to "shift" the gradients from lower to higher input concentrations. Larger variation of the output at large input values, will occur if we work less close to saturation, i.e., decrease the rates of (some of) the reactions. Since gate 1 was already identified as candidate for adjustment, we selected to decrease the (initial) amount of the enzyme GDH.

A new set of data was measured [6], with the concentration of GDH reduced by an order of magnitude, also illustrated in Fig. 4. The results of the data fits are summarized in Table 1. As already emphasized, we are aiming at identifying the regime of *reduced noise amplification*. From this point of view, the results are quite promising: with the use of our simple phenomenological data fitting functions, the noise-amplification measures (see Table 1) came out consistently lower (closer to the optimal) for the modified (improved) network as compared to the original one.

## 4. Conclusion

To recapitulate, we performed analysis and optimization of both a single biochemical logic gate and a three-gate network. We showed that the experimental data for a biochemical gate can be modeled within the rate-equation approach, and the solution of rate equations can be cast in the language of Boolean logic variables. The measured gate response function is used to fit the rate parameters, and the Boolean inputs and output are treated as analog signals in the context of gate-function optimization. We demonstrated both experimentally and theoretically on the example of the AND gate with HRP enzyme that the analog noise generation by the logic gate can be dramatically reduced to achieve virtually no noise amplification.

We also explored a modular approach to analyze performance of an enzymatic network of three AND gates. We developed a methodology which, by avoiding detailed kinetic modeling for each enzymatic gate, allows for selected probes of a complex network to be used for identification and adjustment of parameters for those gates that contribute the most to noise amplification. Our experimental study of such a network offered an illustration of the developed theoretical ideas.

Future work will be focused on studies of larger networks, of systems of interest in applications, and on design of new network elements required to develop the analog/digital paradigm of scalable information processing for biochemical information processing.

We gratefully acknowledge support of our research programs by the US National Science Foundation (grants CCF-0726698, DMR-0706209), and by the Semiconductor Research Corporation (research award 2008-RJ-1839G). G.S. acknowledges the award of the Wallace H. Coulter scholarship at Clarkson University.